\documentstyle[12pt,aaspp4]{article}
\begin{document}
\newcommand{\cen}{\centerline}
\newcommand{\vk}{$\rm km \; s^{-1}\;$}
\title{The Mass-to-Light Ratio of Binary Galaxies}
\author{Mareki Honma\altaffilmark{1}}
\affil{Institute of Astronomy, University of Tokyo, Mitaka, 181-8588, Japan}
\authoremail{honma@milano.mtk.nao.ac.jp}
\altaffiltext{1}{Present Address : VERA project office, National Astronomical Observatory of Japan, Mitaka, 181-8588, Japan}
\slugcomment{submitted to the Astrophysical Journal}

\begin{abstract}
We report on the mass-to-light ratio determination based on a newly selected binary galaxy sample, which includes a large number of pairs whose separations exceed a few hundred kpc.
The probability distributions of the projected separation and the velocity difference have been calculated considering the contamination of optical pairs, and the mass-to-light ratio has been determined based on the maximum likelihood method.
The best estimate of $M/L$ in the B band for 57 pairs is found to be 28 $\sim$ 36 depending on the orbital parameters and the distribution of optical pairs (solar unit, $H_0=50$ km s$^{-1}$ Mpc$^{-1}$).
The best estimate of $M/L$ for 30 pure spiral pairs is found to be 12 $\sim$ 16.
These results are relatively smaller than those obtained in previous studies, but consistent with each other within the errors.
Although the number of pairs with large separation is significantly increased compared to previous samples, $M/L$ does not show any tendency of increase, but found to be almost independent of the separation of pairs beyond 100 kpc.
The constancy of $M/L$ beyond 100 kpc may indicate that the typical halo size of spiral galaxies is less than $\sim 100$ kpc.

\end{abstract}

\section{Introduction}

The mass of a galaxy is a fundamental quantity in understanding its dynamics and structure.
Mass distribution in galaxies has been extensively studied with optical and HI rotation curves.
Several studies revealed that spirals galaxies have flat rotation curves even at the observed outermost points, indicating the existence of extended dark halos (e.g., Sancisi \& van Albada 1987) .
The extent and total mass of dark halos are, however, not understood well and yet to be studied in detail.
For further investigation of extended halos, different approaches are required to trace the mass distribution beyond the HI disk, where rotation curves cannot be measured.

Binary galaxies are useful for the determination of the total mass or mass-to-light ratio ($M/L$) of galaxies, like stellar masses are measured from the motion of binary stars.
Unlike stellar binaries, however, the total mass of individual binary cannot be directly determined because of the long orbital periods.
Instead, statistical treatment is necessary to obtain the average mass or $M/L$ of the sample galaxies.
Many efforts were made to determine the $M/L$ ratio of binary galaxies statistically (e.g., Page 1952; Karachentsev 1974; Turner 1976a, b; Peterson 1979a, b; White 1981; van Moorsel 1987; Schweizer 1987; Chengalur et al.1993; Soares 1996).
In binary galaxy studies, a careful selection of binary galaxies is very important, since the biases in selecting pairs should be corrected for to determine the mass or $M/L$ ratio.
Turner (1976a) proposed well-defined selection criteria based only on the positions and magnitudes of galaxies.
Later investigations (e.g., Peterson 1979a) also made use of similar selection criteria independent of radial velocities, which are so-called {`}velocity-blind{'} pairs.
According to such velocity-blind selection criteria, two galaxies are regarded as a pair if they have no close companion compared to their projected separation.
This {`}velocity-blind{'} selection criterion is simple and convenient for pair selection, but its problem is that the criterion could introduce strong bias toward pairs with small separations; for pairs with wider separations, company galaxies are searched for in a larger region, leading to exclusion of widely-separated pairs with higher probability.
In fact, the average separations of selected pairs in these studies were 50 kpc $\sim$ 100 kpc (see Peterson 1979b).
Since the dark halos could extend beyond this range, it is important to study binary galaxies further based on pairs with wider separations.

The other major problem of the velocity-blind sample is that the sample suffers from the contamination of {`}optical pairs{'}, which consists of two isolated galaxies projected close by chance.
In order to reduce the contamination of optical pairs, it is better to select binary galaxies based not only on the positions but also on the radial velocities.
Fortunately, the number of radial velocity observations is rapidly increasing thanks to recent large-scale redshift surveys.
Moreover, the observational uncertainty has been significantly reduced due to the recent development of observational instruments, which enables us to estimate $M/L$ with better accuracy compared to previous studies.
Therefore, it is interesting to study binary galaxies again by utilizing such huge data.

For these reasons, in this paper we study the mass-to-light ratio of binary galaxies by making use of the database.
The plan of this paper is as follows.
In section 2, we will describe how to select widely-separated pairs effectively, while reducing the contamination of optical pairs.
The selection criteria, and the basic data for selected pairs will be presented in section 2.
In section 3 we will perform maximum-likelihood analysis based on the orbital models of binary galaxies, and determine $M/L$.
We will also consider $M/L$ dependence on galaxies{'} type.
Discussions on the dark halo extent will be given in section 4.

\section{Selection of Pairs}
\subsection{Basic Idea, Selection Criteria, and Sample}

The observable quantities for orbital motion of pairs are the projected separation $r_{\rm p}$, and the radial velocity differences $v_{\rm p}$.
A set of $r_{\rm p}$ and $v_{\rm p}$ can be used to estimate the total mass of a pair through an estimator of mass, for example, $r_{\rm p} v_{\rm p}^2/G$.
However, the mass of galaxies varies by about 3 order of magnitude from dwarf galaxies to giant ellipticals.
A better quantity which represents the mass of galaxies is the mass-to-light ratio, $M/L$.
The mass-to-light ratio of galaxies is expected to vary much less than the mass itself, and hence we focus on the mass-to-light ration of pairs in this paper.

What can be obtained through binary galaxy analysis is the total mass to total light ratio of pairs which is written as $(M_1+M_2)/(L_1+L_2)$, but in the rest of this paper we denote this ratio as $M/L$ for simplicity.
Note that if $M_1/L_1= M_2/L_2$, the total mass-to-light ratio $M/L$ is equal to $M_1/L_1$ and $M_2/L_2$. 
For convenience in $M/L$ estimate, we define the luminosity-corrected separation $R_{\rm p}$ and the luminosity-corrected velocity difference $V_{\rm p}$ as 
\begin{equation}
R_{\rm p} \equiv r_{\rm p}/L^{1/3},
\end{equation} 
\begin{equation}
V_{\rm p} \equiv |v_{\rm p}|/L^{1/3}.
\end{equation}
A combination of $R_{\rm p}$ and $V_{\rm p}$ can give an estimator of the mass-to-light ratio of pairs.
This estimator, which we call the projected mass-to-light ratio, is defined as 
\begin{equation}
(M/L)_{\rm p} \equiv \frac {r_{\rm p} v_{\rm p}^2}{G L} = \frac{R_{\rm p} V_{\rm p}^2}{G}.
\end{equation}
If a bound pair of galaxies are separated so widely that they can be approximated as two point masses, the law of energy conservation gives that
\begin{equation}
\frac{r v^2}{2G L}\le  M/L,
\end{equation}
because the total energy for a bound pair is always negative.
A combination of equation (3) with inequality (4) gives
\begin{equation}
(M/L)_{\rm p} \le 2 (M/L).
\end{equation}
If all pairs are bound and have the same $M/L$, the binary populations in the $R_{\rm p}$-$V_{\rm p}$ phase space lie below the envelope which corresponds to $2(M/L)$.
In practice, the number of pairs is limited and insufficient to see the true envelope corresponding to $2 (M/L)$ in the $R_{\rm p}$-$V_{\rm p}$ space.
Detailed calculations of probability distribution show that pairs are likely to concentrate to small $V_{\rm p}$, and thus smaller $(M/L)_{\rm p}$ due to projection effect(e.g., Noerdlinger 1975).
We will discuss this in later section by calculating the probability distribution based on the Monte-Carlo simulation.
In any case, the pair distribution in the $R_{\rm p}$-$V_{\rm p}$ phase space can be used for testing whether or not bound pairs are efficiently selected: while bound pairs are likely to have small $V_{\rm p}$, optical pairs could have extremely high $V_{\rm p}$ and $(M/L)_{\rm p}$.

Here we describe the selection criteria for pairs.
For convenience, we define the total luminosity of a pair normalized with $10^{10}L_\odot$ in the B band as $L_{10}\equiv (L_1+L_2)/(10^{10}L_\odot)$.
Note that this luminosity roughly corresponds to that of the Milky Way Galaxy.
We re-define $R_{\rm p}$ and $V_{\rm p}$ normalizing with luminosity $L_{10}$ as,
\begin{equation}
R_{\rm p} = \frac{r_{\rm p}}{L_{10}^{1/3}} \;{\rm (kpc)}, 
\end{equation}
and
\begin{equation}
V_{\rm p} = \frac{|v_{\rm p}|}{L_{10}^{1/3}} \;{\rm (km\; s^{-1})}.
\end{equation}

As the first step of pair selection, we have to select a pair of galaxies that are relatively close to each other both in the sky plane and in the redshift space so that they are likely to be bound.
Since the average separations in previous studies were around 100 kpc, and since we are interested in widely-separated pairs, we set the maximum projected separation of a pair in the sky plane as 
$$
1: R_{\rm p} \le 400 \;{\rm kpc}.
$$

The maximum velocity difference must also be large enough to include pairs orbiting around each other at high velocity.
Since galaxies as luminous as $10^{10}L_\odot$ have rotation velocity of $\sim$ 200 km s$^{-1}$, the velocity difference of a pair could be as high as a few hundred km s$^{-1}$.
Hence, we set the maximum velocity difference of pairs as
$$
2: V_{\rm p} \le 400 \;{\rm km\; s^{-1}}.
$$
Note that the radial velocity difference is usually quite small compared to the true velocity difference due to projection effect (Noerdlinger 1975).
Thus, physical pairs are unlikely to have larger radial velocity differences than the maximum value given above.

The observational data set cannot be complete to faint galaxies, and hence we limit the application of our analysis to sufficiently bright galaxies.
Since nearby galaxies are cataloged almost completely down to 15.5 magnitude (e.g, de Vaucouleurs et al. 1991), we set a criterion for the B band magnitude as
$$
3: m_1, m_2 \le 15.0 \;{\rm mag},
$$
and also set an upper limit for the total magnitude of a pair in the B band, $m_{1+2}$, as
$$
4: m_{1+2} \le 13.5 \;{\rm mag}$$

In order for selected pairs to be likely to be bound, pair galaxies must be isolated well.
We regarded two galaxies as a pair if all of its companion galaxies brighter than $m_{1+2} + 2.0 \;{\rm mag}$ satisfy both
$$
5: \frac{r_i}{L_{10}^{1/3}} \ge a\; 400 \; {\rm kpc},$$
and
$$
6: \frac{|v_i|}{L_{10}^{1/3}} \ge b\; 400 \; {\rm km s^{-1}}.$$
Here $r_i$,and $v_i$ are the projected separation and the radial velocity difference of $i$th companion galaxy with respect to the luminosity center of the pair.
Note that we set the lower limit of the total magnitude $m_{1+2}$ to be 13.5 mag (criterion 4).
The faintest galaxies that should be considered is, hence, at 15.5 mag, to which magnitude galaxies are cataloged almost completely.

Parameters $a$ and $b$ determine the volume for companion search, and so determine the degree of isolation.
Note that the volume depends only on the total luminosity of a pair, $L_{10}$, but independent of the separation $R_{\rm p}$ of a pair.
Therefore, as far as pairs with the same luminosity are concerned, companions are searched for in the same volume, and thus the criterion does not introduce bias toward pairs with small separations.
We set $b=1.5$ throughout this paper, but tested three values of $a$ (1.5, 2.0, and 2.5) to seek a value for effective selection of bound pairs.

We applied criteria described above to the sample of galaxies that we compiled for this study using NED (NASA Extra-galactic Database).
The sample consists of bright nearby galaxies with redshift less than 4,500 $\rm km \; s^{-1}\;$.
The upper limit for redshift is introduced because the number of bight pairs which satisfy the criteria 3 and 4 becomes small at large redshift.
The data for positions, heliocentric velocities, and the B band magnitudes were mainly taken from NED, and supplied with RC3 catalog (de Vaucouleurs et al.1991).
The distances to the pairs are obtained using the redshift of the luminosity center ($H_0=$ 50 $\rm km \; s^{-1}\;$ is assumed).
In order to avoid the error in distance due to the local deviation from Hubble flow, galaxies with redshift smaller than 1,000 $\rm km \; s^{-1}\;$ are excluded from the sample.
Galaxies in clusters and close to clusters may deviate from the Hubble flow even beyond the redshift of 1000 $\rm km \; s^{-1}\;$, but this effect is expected to be small because the pairs selected with criteria 5 and 6 are likely to be field binary galaxies.
Galaxies with $b \le 20^\circ$ are also excluded from the sample since objects at low galactic latitude may be significantly obscured by galactic extinction
The sample we compiled consists of 6475 galaxies with magnitude brighter than 15.5 mag and redshift between 1000 km s$^{-1}$ and 4500 km s$^{-1}$.
The uncertainty in the magnitude is typically 0.2 mag., which leads to the uncertainty in $L$ of 20 \%.
The corrections for intrinsic absorption and galactic extinction were made according to de Vaucouleurs et al.(1991).

The sample is, of course, incomplete in terms of redshift because redshift measurements were not made for all galaxies.
This incompleteness leads to possible mis-identification of pairs, if the criteria described above are applied only to a sample of redshift-know galaxies.
To correct for the effect of redshift incompleteness, primary binary candidates are at first searched in the sample of redshift-known galaxies, and then a redshift-blind search was performed for the primary binary candidates.
If there is any redshift-unknown companion which is brighter than $m_{1+2} + 2.0$ mag and is so close to the pair that the criterion 5 is violated, the pair was rejected from the binary candidates.
About 30 \% of pairs in the primary binary candidates were rejected through this procedure.

The sample of binary galaxies after the correction for the redshift incompleteness still contains some pairs that are not appropriate for this study.
For instance, the basic data for the analysis, such as $v_{\rm p}$, $m_1$, and $m_2$ could be quite uncertain for some pairs.
In particular, the uncertainty in the radial velocity is crucial for $M/L$ determination, as the $M/L$ estimator depends on $V_{\rm p}^2$.
Therefore, if redshift uncertainty is not reported for any of the two galaxies of pair, the pair is excluded. 
This process reduced the number of pairs by 7\%.
If the magnitude uncertainty and the absorption-corrected magnitude are not available, the pair is also excluded, and in this process 7\% of primary binary candidates were rejected.
Moreover, a galaxy could appear in the binary sample twice or more with different partner.
This can happen if one of pair galaxies is a bright galaxy like the cD galaxy and it has several companion galaxies around it.
In this case, however, these galaxies should be regarded as cluster or group rather than binary galaxies.
Therefore, we also excluded possible clusters or groups of galaxies that appear in the binary sample twice or more.
We found only two possible groups in the primary binary candidates.

\subsection{Results}

Figures 1 show the distribution of thus selected binary galaxies in the $R_{\rm p}$-$V_{\rm p}$ phase space.
Two cases for the isolation parameter, $a$=1.5 and 2.5, are shown.
The number of selected pairs is 109 and 57, respectively.
In the case of $a$=1.5, the pairs in the $R_{\rm p}$-$V_{\rm p}$ space shows only weak concentration toward small $V_{\rm p}$, and a large number of galaxies have high $(M/L)_{\rm p}$ exceeding a few hundred $M_\odot/L_\odot$.
Even if their true $M/L$ is a few hundred, it is unlikely that so many galaxies appear to have so large $(M/L)_{\rm p}$ in the projected phase space, as $(M/L)_{\rm p}$ is expected to be significantly smaller than the true $M/L$ due to projection effect (Noerdlinger 1975; see also Section 3 of the present paper).
This indicates that they are probably optical pairs, and that the degree of isolation is not strong enough to select bound pairs effectively.
On the other hand, for $a=2.5$, the concentration of pairs to $V_{\rm p}$=0 is much more clear than for $a=1.5$.
Most of 57 pairs are distributed below $(M/L)_{\rm p}$ of 20 in solar unit, and there are only few galaxies that have high $(M/L)_{\rm p}$.
This correlation between $R_{\rm p}$ and $V_{\rm p}$ are naturally explained if the separation of pairs are larger than the extent of halos so that the pairs can be approximated as point masses (but note that even in case of extended hale such an envelope would appear in the projected phase-space like point-mass cases; see Soares 1990).
In the rest of this paper, we use the binary galaxies sample selected with $a=1.5$ and 2.5 for $M/L$ determination.
We call the sample selected with $a=2.5$ as sample I, and the one selected with $a=1.5$ as sample II.
Table 1 summarizes the basic data for 57 pairs in sample I.

\section{$M/L$ determination}

In this section, we estimate the $M/L$ ratio of the sample pairs selected above.
We construct the orbital models for physical pairs, and calculate the probability distribution of pairs in the $R_{\rm p}$-$V_{\rm p}$ phase space considering the contamination of optical pairs.
Then, we compare the models with the observational data, and determine the $M/L$ ratio based-on maximum-likelihood analysis.

\subsection{Distribution of Bound Pairs}

First we construct models for orbital populations of binaries.
For simplicity, binary galaxies are treated as point masses in the following analysis.
As can be seen in figure 1b, pairs show strong concentration toward small $V_{\rm p}$ in the $R_{\rm p}$-$V_{\rm p}$ space, which is just expected from the point-mass assumption.
Further tests for validity of the assumption will be made in the next section.

An ensemble of well-mixed binary population satisfy the Jeans equation (Binney and Tremaine 1987),
\begin{equation}
\frac{d (\nu \overline{v_{\rm r}^2})}{d r} + \frac{2\nu \beta \overline{v_{\rm r}^2}}{r}=-\frac{GM}{r^2}\nu.
\end{equation}
Here $\nu$ denotes the separation distribution of pairs, and $\beta$ is the anisotropy parameter defined as
\begin{equation}
\beta = 1 - \frac{\overline{v_\theta^2}}{\overline{v_{\rm r}^2}},
\end{equation}
where $\overline{v_\theta}$ and $\overline{v_{\rm r}}$ denote each component of velocity ellipsoids.
Note that $\beta = -\infty$ for circular orbits, $\beta = 0$ for isotropic orbits, and $\beta = 1$ for radial orbits.
We may rewrite equation (8) by normalizing with luminosity as
\begin{equation}
\frac{d (\nu \overline{V_r^2})}{d R} + \frac{2\nu \beta \overline{V_r^2}}{R}=-\frac{G M}{R^2 L}\nu,
\end{equation}
where $R=r/L^{1/3}$ and $\overline{V_r^2}=\overline{v_r^2}/L^{2/3}$.
Note that they are true separation and velocity, but not projected ones.

For the separation distribution $\nu$, we also assume a power law with an inner cutoff radius $r_{\rm min}$ as,
\begin{equation}
\nu(R) \propto R^{-\gamma} \;\;\;\;\; {\rm for} \;\;\;\; R \ge R_{\rm min}.
\end{equation}
We introduced the cutoff radius because galaxies have finite sizes, and pairs that are too close are not likely to exist.
The model used here is, therefore, not exactly a scale-free model (cf. White 1981).

In order to model the distribution of pairs, one should choose suitable values for parameters $\beta$, $\gamma$, and $R_{\rm min}$.
The parameters related to the separation distribution are obtained directly from observed separation distribution, because the probability distribution for projection effect can be written analytically as
\begin{equation}
p [R_{\rm p} | R] = \frac {2 R_{\rm p}}{\pi R(R^2 - R_{\rm p}^2)^{1/2}} \;\;\;\;\;\;\;\; ({\rm for\;\;\;} R_{\rm p} \le R).
\end{equation}

We compared the separation distribution of observed and model pairs, and obtained the best-fit values $\gamma = 2.6$ and $R_{\rm min}= 10$ kpc.
In the rest of this paper, we adopt these best-fit values for $\gamma$ and $R_{\rm min}$, but we note the results are not sensitive to changes in the assumed values.
Once the separation distribution is obtained, the distribution of the velocity difference is obtained by solving the Jeans equation [eq.(8)]. 
Then, one can calculate the probability distribution of pairs in the $R_{\rm p}$-$V_{\rm p}$ phase space, $p_{\rm bin} [R_{\rm p},V_{\rm p}|(M/L)]$, by taking the projection effect into consideration.

\subsection{Selection Bias and Contamination of Optical Pair}

In addition to the orbital models for true pairs, here we consider the selection effect for pairs and the contamination of optical pairs.
As described in the previous section, the isolation criteria are independent of the separation or velocity difference of pairs, and hence, the sample is free from biases both for the separation in the sky plane, and for the separation along the line of sight.
The isolation criteria, however, may cause possible exclusion of true pairs due to chance projection of another companion galaxy that are not physically related to the pair.
According to the isolation criteria, the maximum velocity difference of a pair is $\sim 400$ $\rm km \; s^{-1}\;$ for galaxies with $L\sim 10^{10} L_\odot$.
Therefore, a system of a true pair plus any foreground or background galaxy at distant within 8 Mpc from the pair cannot be regarded as a true pair because the criterion 6 is violated ($H_0 =50$ km s$^{-1}$ Mpc$^{-1}$ assumed).
Unfortunately, there is no way to determine whether the observed velocity difference of two galaxies is due to Hubble flow or due to binary orbital motion, and so this kind of exclusion of true pairs is unavoidable.
Furthermore, two galaxies which are separated well along the line of sight and are not physically associated could be regarded as a pair because of misidentification of the redshift as binary orbital motion.
For these reasons, the sample selected in the present paper is far from perfect but likely to contain unphysical pairs which would lead to wrong estimates of $M/L$.
Therefore, the exclusion of true pairs and the contamination of optical pairs must be taken into consideration for $M/L$ determination.

Fortunately, the possibility of true-pair exclusion is independent of $R_{\rm p}$ or $V_{\rm p}$, as the selection criteria do not depend on them.
Hence, the probability distribution of pairs in $R_{\rm p}$-$V_{\rm p}$ space, which is to be compared with the observed pairs, can be expressed as
\begin{equation}
p[R_{\rm p}, V_{\rm p}|M/L,f] = f p_{\rm bin}[R_{\rm p}, V_{\rm p}|(M/L)] + (1-f) p_{\rm opt}[R_{\rm p}, V_{\rm p}],
\end{equation}
where $f$ is a constant corresponding to the fraction of true pairs out of observed pairs.
Clearly the first term on the right side expresses the contribution of true binaries, and the second term describes the contamination from optical pairs.
Probabilities $p$, $p_{\rm bin}$, and $p_{\rm opt}$ are normalized so that
\begin{equation}
\int \int p\; dR_{\rm p}dV_{\rm p} = \int \int p_{\rm bin}\; dR_{\rm p}dV_{\rm p} = \int \int p_{\rm opt}\; dR_{\rm p}dV_{\rm p} = 1,
\end{equation}
where the integrations are performed from 0 to 400 kpc for $R_{\rm p}$, and from 0 to 400 $\rm km \; s^{-1}\;$ for $V_{\rm p}$.

The possibility of mis-identification of optical pairs is proportional to the number density of galaxies.
It is generally known that the distribution of galaxies in the Universe is not uniform but shows strong clustering, which is usually described in terms of the two-point correlation function (e.g. Peebles 1993).
With this function the probability distribution of optical pairs in the $R_{\rm p}$-$V_{\rm p}$ phase space can be written as
\begin{equation}
p_{\rm opt}[R_{\rm p},V_{\rm p}] \;dR_{\rm p} dV_{\rm p}\propto \left[1+\xi(r)\right] R_{\rm p} \;dR_{\rm p} dV_{\rm p},
\end{equation}
where $\xi(r)$ is the two-point correlation function, and this is usually written in the form of
\begin{equation}
\xi(r) = \left( \frac{r_0}{r} \right)^{q}.
\end{equation}
The two-point correlation function is determined well in the scale of 10 Mpc but less certain in the scale of 1 Mpc.
Hence in the following analysis we consider two cases, no clustering case with $q=0$ and clustering case with $q=1.8$ and $r_0=10$ Mpc (Peebles 1993), and see how the $M/L$ estimates depend on the clustering effect.
The separation $r$ can be calculated from the projected separation and the velocity difference by assuming the Hubble constant of 50 km s$^{-1}$ Mpc$^{-1}$.
Note that in any case the probability distribution of optical pairs in the $R_{\rm p}$-$V_{\rm p}$ space is independent of the $M/L$ ratio of galaxies.

\subsection{$M/L$ Determination}

To evaluate the mass-to-light ratio and the fraction of true pairs we make use of the maximum-likelihood method for $M/L$ and $f$.
The probability for finding a pair at $R_{\rm p}$ and $V_{\rm p}$ in the projected phase space is proportional to $p[R_{\rm p},V_{\rm p}|M/L,f]$, and hence the logarithmic likelihood for finding all observed pairs at their observed positions in the projected phase space is expressed as the summation of the probability for finding each pair at its position.
Therefore, the logarithmic likelihood of $M/L$ for observed pairs can be written as
\begin{equation}
\log {\cal L} (M/L,f) = \sum n(R_{\rm p}, V_{\rm p}) \; \log p\,[R_{\rm p}, V_{\rm p}|M/L,f],
\end{equation}
where $n(R_{\rm p}, V_{\rm p})$ denotes the observed number of pairs having $R_{\rm p}$ and $V_{\rm p}$, and the summation is done over the whole projected phase space ($R_{\rm p}$ less than 400 kpc and $V_{\rm p}$ less than 400 km/s).
Evidently $n(R_{\rm p}, V_{\rm p})$ is integral as long as the values of $R_{\rm p}$ and $V_{\rm p}$ are determined with sufficient accuracy.
However, for the pairs we consider here $R_{\rm p}$ and $V_{\rm p}$ have uncertainties, and the uncertainty in $V_{\rm p}$ is particularly crucial for $M/L$ determination because the $M/L$ estimator depends on $V_{\rm p}^2$.
Therefore, we treated each observed pair as a Gaussian distribution spread in the direction of $V_{\rm p}$, and then $n(R_{\rm p}, V_{\rm p})$ is given as
\begin{equation}
n(R_{\rm p}, V_{\rm p})\; dR_{\rm p} dV_{\rm p} \propto \sum_i g_i \; dR_{\rm p} dV_{\rm p},
\end{equation}
where 
\begin{equation}
g_i = (2\pi \sigma_i)^{-1/2} \exp \left(\frac{-(V_{\rm p}-V_i)^2}{2\sigma_i^2}\right).
\end{equation}
Here $V_i$ and $\sigma_i$ denote the observed $V_{\rm p}$, and the uncertainty for $V_{\rm p}$ for $i$th pair, respectively.
Note that $n$ is normalized so that $\int \int n \; dR_{\rm p} dV_{\rm p}= N_{\rm tot}$, where $N_{\rm tot}$ is the total number of pairs.

Since the probability distribution of physical pair in the $R_{\rm p}$-$V_{\rm p}$ phase space, $p_{\rm bin}[R_{\rm p}, V_{\rm p} | (M/L)]$, cannot be expressed analytically due to the projection effect, we performed Monte-Carlo simulations to evaluate $p_{\rm bin}$.
The distribution of one million pairs in the projected phase space were simulated assuming random orientation of orbital planes with respect to the line of sight, and then the logarithmic likelihood (equation [16]) was calculated in the parameter space of $f$ and $M/L$.


Figure 2 shows the likelihood contours for sample I (57 pairs) for the case of $q=0$ (no clustering for optical pairs).
The thick lines are for $\beta =0$ (isotropic orbit) and dotted lines are for $\beta = -\infty$ (circular orbit).
The figures show that the M/L estimates are not affected strongly by the assumed orbital parameters.
The best estimates of $M/L$ are $35_{-5}^{+7}$ for $\beta = 0$ and $28_{-3}^{+5}$ for $\beta = -\infty$, with true binary fraction $f$ of $0.88_{-0.1}^{+0.07}$ for both cases (the error bars denote the 68 \% confidence level).
The results for the true binary fraction $f$ indicates that most of pairs in sample I are likely to be bound.
The expected number of optical pair is about 7 out of 57 pairs, which is comparable to the number of pairs which appear in Figure 1b above the envelope corresponding to $(M/L)_{\rm p}$ of 20.

On the other hand, figure 3 shows the likelihood contours for sample I like figure 2, but for $q=1.8$ (clustering for optical pairs).
The contours for two orbital models are shown, and again one can see that the weak dependence of $M/L$ on the orbital parameter.
However, the true pair fractions $f$ are quite different from those for no clustering cases, as we obtained $f=0.71_{-0.15}^{+0.14}$ ($\beta = 0$) and $f=0.73_{-0.15}^{+0.14}$ ($\beta = -\infty$) for clustering case.
This is because the expected number of optical pairs with small $V_{\rm p}$ is much larger than that for no clustering case, and hence more number of galaxies with small $V_{\rm p}$ are regarded as optical pairs.
However, the best estimates of $M/L$ are $36_{-4}^{+8}$ for $\beta = 0$ and $30_{-3}^{+6}$ for $\beta = -\infty$, which are fairly close to those for no clustering cases.

In figure 2 and 3 the best estimates of $M/L$ change little depending on $f$.
This is because the $M/L$ is essentially determined by the pairs which lie below the envelope in figure 1b: in the most range of $f$ (e.g. $f$ less than 0.95) the pairs far above the envelope are likely to be optical pairs, and have little effect on the $M/L$ determination.
On the other hand, if $f$ is set to be almost unity, the most likely $M/L$ could become as high as 100 to explain the pairs with high $(M/L)_{\rm p}$ without optical pairs.
However, the likelihood for finding $f$ of almost unity is very small compared to that for $f$ between 0.6 to 0.9, for in that case the concentration of pairs to small $(M/L)_{\rm p}$ seen figure 1 cannot be explained at all.

In order to test whether $M/L$ and $f$ obtained above can reproduce the distribution of observed pairs, we simulated distribution of model pairs in the $R_{\rm p}$-$V_{\rm p}$ phase space using the best-fit parameters.
Figure 4 shows the simulated distribution of 57 modeled pairs with $f=0.88$ and $M/L=35$ assuming isotropic orbits for true pairs and no clustering for optical pairs.
The figure resembles well the observed distribution in figure 1b in many aspects; small fraction of pairs with extremely high $(M/L)_{\rm p}$, concentration of pairs to small $V_{\rm p}$, and the envelope corresponding to $(M/L)_{\rm p}\sim 20$.
This simulation confirms the validity of the results.

In order to see if these results depend strongly on the sample we used, we also performed the same analysis for sample II (109 pairs), which are selected with weaker isolation criterion ($a=1.5$).
As seen in Figure 1a, this sample is likely to contain more number of optical pairs than sample I due to the weak selection criterion.
In fact, the resultant value of $f$ is 0.80$_{-0.08}^{+0.07}$ for no clustering case ($q=0$) and $0.62_{-0.11}^{+0.10}$ for clustering case ($q=1.8$), with $\beta=0$ for both cases.
However, the best estimates $M/L$ are $35_{-3}^{+5}$ ($\beta=0$) and $28_{-2}^{+5}$ ($\beta=-\infty$) for no clustering cases, and $36_{-3}^{+12}$ ($\beta=0$) and $30_{-3}^{+8}$ ($\beta=-\infty$) for clustering cases.
These results remarkably agree with those for sample I, indicating that the results does not depend strongly on the samples.
The results for sample II as well as those for sample I are listed in Table 2.

However, we would like to note that the results might be changed if we take the other limit of $\beta$; $\beta=1$ corresponding to radial orbits.
In this case the best $M/L$ for sample I was found to be $42_{-7}^{+34}$ ($q=0$), and so the $M/L$ would exceed 50 within 68 \% confidence level.
Yet the assumption of $\beta=1$ seems too radical, because in this case the pairs suffer from direct encounters that will probably lead them to mergers.
In order for bound pairs to survive for many orbital periods, their orbits must be elliptical at least to some degree, and so $\beta$ cannot be too close to unity.
On the other hand, perfectly circular orbits ($\beta =-\infty$), are also unlikely because this requires fine tuning of orbital parameters.
Therefore, the results with $\beta=0$ presumably represent best the mass-to-light ratio of true pairs.

\subsection{$M/L$ for pure spiral pairs}

In the $M/L$ determination above, we did not consider the variation of $M/L$ among the sample galaxies.
The $M/L$ of galaxies are, however, usually considered to vary depending on the type of galaxies.
Indeed, previous binary galaxy studies claimed larger $M/L$ for ellipticals than that of spirals (e.g., Schweizer 1987).
Therefore, it is interesting to study the $M/L$ ratio of galaxies for a specific type.

The 57 pairs in sample I consist of spirals, S0s, ellipticals and some others such as peculiars.
The dominant type among them is spirals, which occupies a fraction of 70\% in sample I, and hence we try to estimate $M/L$ for spiral galaxies.
In particular we concentrate on {`}pure{'} spiral pairs which consists of two spiral galaxies later than Sa, because a pair of a spiral and an S0 or elliptical do not necessarily reflect the $M/L$ of spiral galaxies.
We selected 30 pure spiral pairs out of 57 pairs in sample I (hereafter sample III), and performed the same analysis described above.
The likelihood contours for no clustering case ($q=0$) and for clustering case ($q=1.8$) are plotted in figure 5 and 6, respectively.
Most likely value for no clustering case is found to be $15_{-3}^{+5}$ for $\beta = 0$, and $12_{-3}^{+4}$ for $\beta=-\infty$, which are compared to the results for the 57 pairs with mixed types, 35 for $\beta=0$ or 28 for $\beta=-\infty$.
The best estimates of $M/L$ for clustering case are similar to those for no clustering case whereas the true pair fraction $f$ is relatively smaller (see table 2).
In both cases the difference in $M/L$ between sample I and III is significant, being above the 3$\sigma$ level.
Therefore, we conclude the difference is real, and that $M/L$ for spirals are smaller to ellipticals or S0s.
This is consistent with previous studies of binary galaxies, although the $M/L$ obtained here are somewhat smaller than those from previous studies.

\section{Discussion}
\subsection{$M/L$ Dependence on Separation}

In the previous sections, pairs are approximated as two point masses.
However, real galaxies have finite size, and it could be as much as 100 kpc if dark halos extend well beyond the optical disks.
In this case, the approximation of point masses is not valid, and $M/L$ obtained above could be underestimation, particularly for pairs with small separations.
Here we investigate the dependency of $M/L$ on the separation of pairs, and test if the assumption of point masses is reasonable for the present samples.

We divided 57 pairs in sample I into 3 subgroups depending on the separation.
The three subgroups consist of 27 pairs with $0 <R_{\rm p}< 100$ kpc, 12 with $100 <R_{\rm p}< 200$ kpc, and 18 with $200 <R_{\rm p}$ kpc, respectively.
The $M/L$ ratios for three subsample were obtained in the same manner described in the previous section.
Assuming $\beta = 0$ and $q=0$, we obtained the best estimates of $M/L$ with $1\sigma$ errors to be  $36_{-5}^{+10}$, $37_{-17}^{+31}$, and $25_{-13}^{+29}$, respectively.
The error bars are increased compared to the results in Section 3 because the number of galaxies in a sample is reduced.
figure 7 shows the $M/L$ dependence on the mean separations of pairs.
The figure demonstrates that the $M/L$ ratio is almost constant, and that the variations are within the $1\sigma$ error bars.
If the density distribution of dark halo is proportional to $r^{-2}$ at a large radius, the $M/L$ increases linearly with radius, and if it is proportional to $r^{-3}$ as suggested by recent simulations (Navarro et al.1996), $M/L$ increases with radius logarithmically.
However, figure 7 shows no tendency of increasing $M/L$ at large radii.
Therefore, the halos of galaxies as luminous as the Milky Way Galaxies may be truncated within 100 kpc.
The indication of constant $M/L$ beyond 100 kpc is consistent with previous studies of binary galaxies.
According to Peterson (1979b), the $M/L$ ratio is gradually increasing with radius at $R<100$ kpc, but remains constant beyond it.
Schweizer (1987) also showed that $M/L$ does not increase beyond 100 kpc.

The $M/L$ estimate for spiral galaxies is also consistent with previous studies.
Schweizer (1987) obtained $M/L$ of $21\pm 5$ (V band, absorption corrected) with the sample whose mean separation is about 90 kpc.
Peterson (1979b) obtained spiral galaxies{'} $M/L$ of $35\pm 13$ ($H_0 = 50$ km s$^{-1}$ Mpc$^{-1}$) based on 39 pairs with their mean separation of 110 kpc, and Turner (1976b) also obtained $M/L$ for spiral galaxies of $\sim$ 35.
Note that in the 70's the correction for the galactic and internal absorption were not usually made, and this partly explains smaller $M/L$ in the present paper.
The mean of absorption correction in our sample galaxies is about 0.4 mag, which reduces $M/L$ by $\sim$ 30 \%.
Therefore, if similar amount of absorption correction were made, the studies by Peterson (1979b) or Turner (1976b) would give $M/L$ of 23$\sim$27, which are close to our results of $M/L$.
Although the $M/L$ obtained in the present paper is not significantly different from previous studies, we would like to emphasize that the mean of absolute separations (not the luminosity corrected separation $R_{\rm p}$) of sample III in this paper is $\sim$ 206 kpc, which is almost twice of those in previous studies.
Nevertheless, the $M/L$ ratio does not show any tendency of increase with increasing the separation when compared with previous studies.

\subsection{Dark Halo Extent of Spiral Galaxies}

Since the mass of spiral galaxies{'} optical disk can be estimated from rotation curves, we can compare the total $M/L$ with optical disk $M/L$, and discuss the extent of dark halos of spiral galaxies.
We define $(M/L)_{R25}$ as the ratio of the enclosed mass within $R_{25}$ to the total luminosity, where $R_{25}$ is a radius at which the surface brightness becomes 25 mag per arcsec$^2$.
The enclosed mass can be estimated from the HI velocity line width $W_{\rm HI}$.
Assuming that $W_{\rm HI}$ corresponds to twice of the rotation velocity, we obtain 
\begin{equation}
(M/L)_{R25} = \frac{R_{25} W_{\rm HI}^2}{4 G L}.
\end{equation}
The central surface luminosity of spiral galaxies is constant, about 22 mag per arcsec$^2$ (Freeman 1970).
If this applies to the spiral galaxies in the binary sample, $R_{25}$ corresponds to about 3 times disk scale length $d$, and $(M/L)_{R25}$ roughly approximates the mass-to-right ratio of the disk.
We calculated $(M/L)_{R25}$ of spiral galaxies in sample I for which $R_{25}$ and $W_{\rm HI}$ are available.
$R_{25}$ and $W_{\rm HI}$ were taken from de Vaucouleurs et al.(1991), and Huchtmeier and Richter (1989), respectively.
The values of $(M/L)_{R25}$ range from 2 to 18 with the average of 7.
Note that this $M/L$ is consistent with the previous studies for disk $M/L$; for example, Faber \& Gallagher (1979) obtained the $M/L$ of about 5 ($H_0=50$ km s$^{-1}$ Mpc$^{-1}$).
This $M/L$ is compared to the total $M/L$ obtained in the section 3, $M/L$ of 12 $\sim$ 16.
The total $M/L$ is somewhat larger than the disk $M/L$, and the difference is almost $3\sigma$ level (see figures 5 and 6).
This difference is of course due to the dark halo, and this indicates that under the maximum disk assumption the contributions of dark halo and the optical disk to the total mass of galaxies are comparable.
However, the assumption of maximum disk is still controversy.
If the disk mass is smaller than that indicated by the maximum rotation velocity within the optical disk, the dark halo could dominantly contribute to the total mass.

We can also estimate the extent of dark halos by comparing the total $M/L$ with disk $M/L$.
If a flat rotation curve is assumed, the value of $M/L$ increases linearly with radius.
In this case, the resultant $M/L$ of 15 implies that the typical halo extends $15/7 \approx 2$ times $R_{25}$.
If we adopt disk $M/L$ of 5 according to Faber \& Gallagher (1979), the halo extent is $15/5 \approx 3$ times $R_{25}$.
Therefore, the typical dark halo size maybe about 6 to 9 times disk scale length, if $R_{25}$ is $\sim 3$ times disk scale length $d$.
This is to be compared with the size of optical disk, which is about $4\sim 5$ times disk scale length (van der Kruit \& Searle 1981).
Hence, if the rotation curve is perfectly flat out to the radius at which the halo mass distribution is truncated, the halo size may be about 2 times larger than optical disks.
This is, of course, not exactly true when the rotation curve is not completely flat, which is preferred by the recent simulations.
In this case, the halo size may be somewhat lager, but it cannot exceed a few hundred kpc as indicated by the $M/L$ constancy.

The halo size indicated here is somewhat smaller than those in previous studies, but quite consistent with recent investigations.
For instance, a number of declining rotation curves, which may be fitted even with a Keplerian, are found recently (e.g., J\"ors\"ater \& van Moorsel 1995; Olling 1996; Honma \& Sofue 1996).
Honma \& Sofue (1997) showed that such declining rotation curves are not uncommon by considering the observational uncertainty.
These rotation curves studies are generally based on the observation of HI, which are usually observed out to $5\sim 10$ times disk-scale length.
Therefore, the fact that the declining part of rotation curves were found is consistent with the present results.

\acknowledgments

We are grateful to Y. Sofue for his supervision, and to Y. Tutui and J. Koda for fruitful discussion.
This work was financially supported by the Japan Society for the Promotion of Science.

\clearpage

\section*{figure captions}
\def\r{\hangindent=1pc \noindent}

\r Figure 1. Distribution of selected pairs in the $R_{\rm p}$-$V_{\rm p}$ phase space.
Figure 1a is for the pairs selected with $a=1.5$, and 1b is for the pairs selected with $a=2.5$.
The dotted curve in figure 1b corresponds to the constant $(M/L)_{\rm p}$ of 20.

\r Figure 2. Likelihood contours in the parameter space of $M/L$ and $f$ for sample I.
As for the optical pair distribution no clustering is assumed ($q=0$).
Solid lines are for $\beta=0$ (isotropic orbits), and dotted lines are for $\beta=-\infty$ (circular orbits).
Three contours for each case correspond to 68, 95 and 99\% level, and the crosses denote the peak of the likelihood.

\r Figure 3. Likelihood contours same to figure 2, but for $q=1.8$ (clustering case for optical pairs).

\r Figure 4. Simulated distribution of 57 pairs which are to be compared with figure 1b.
The distribution is calculated with the best fit parameters obtained in the Section 3.
The dotted curve corresponds to $(M/L)_{\rm p}$ of 20.

\r Figure 5. Likelihood contours same to figure 2 (no clustering case), but for sample III (30 pure spiral pairs).

\r Figure 6. Likelihood contours same to figure 3 (clustering case), but for sample III (30 pure spiral pairs).

\r Figure 7. $M/L$ for three subgroups against the mean separation (see text for the sample).
The error bar for $R$ denotes the $1\sigma$ deviation of pairs in the sample.

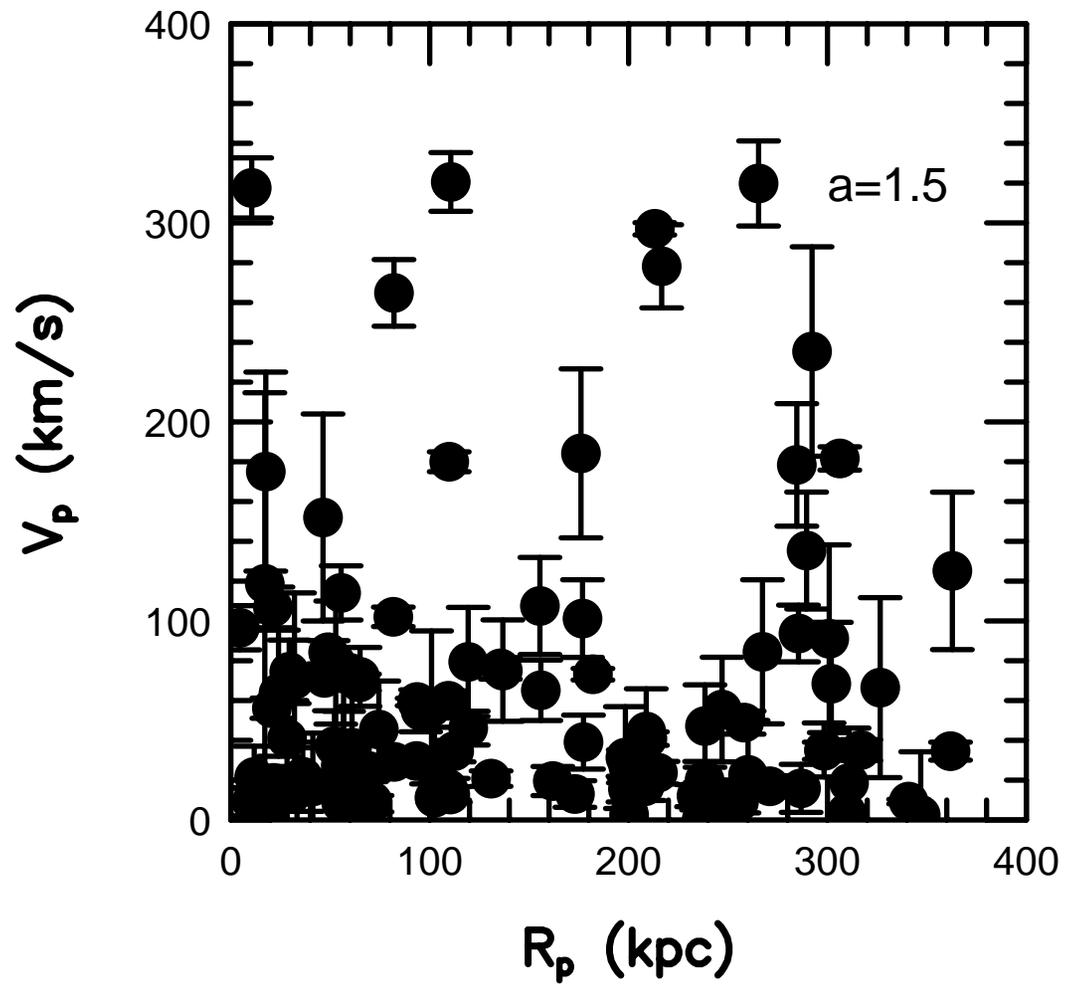
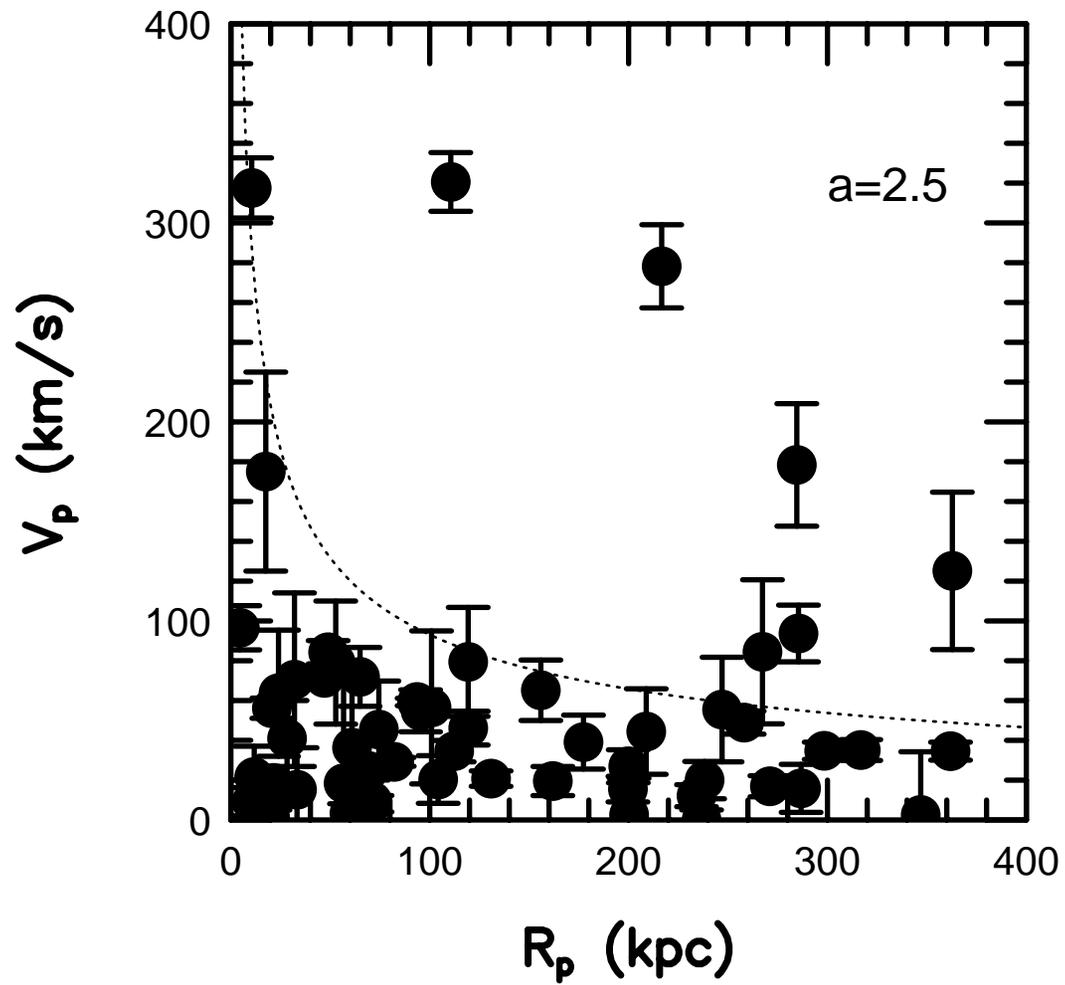
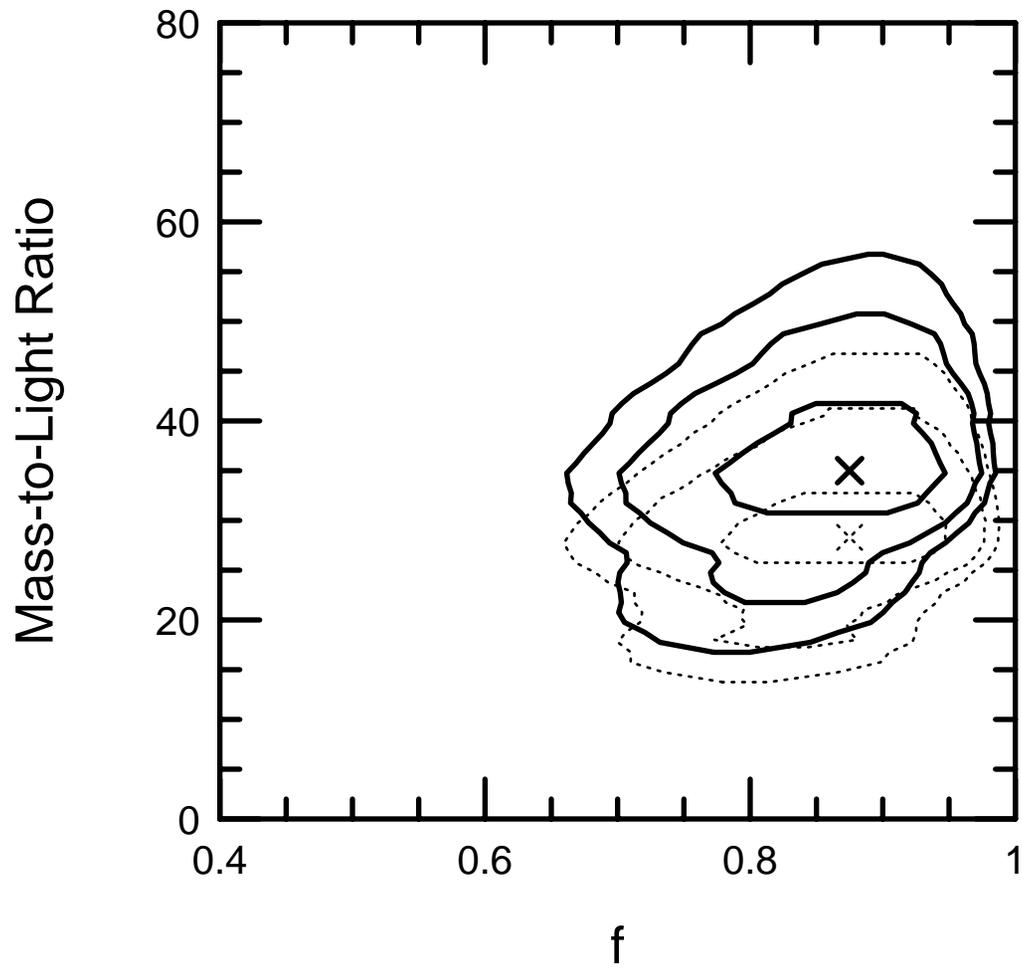
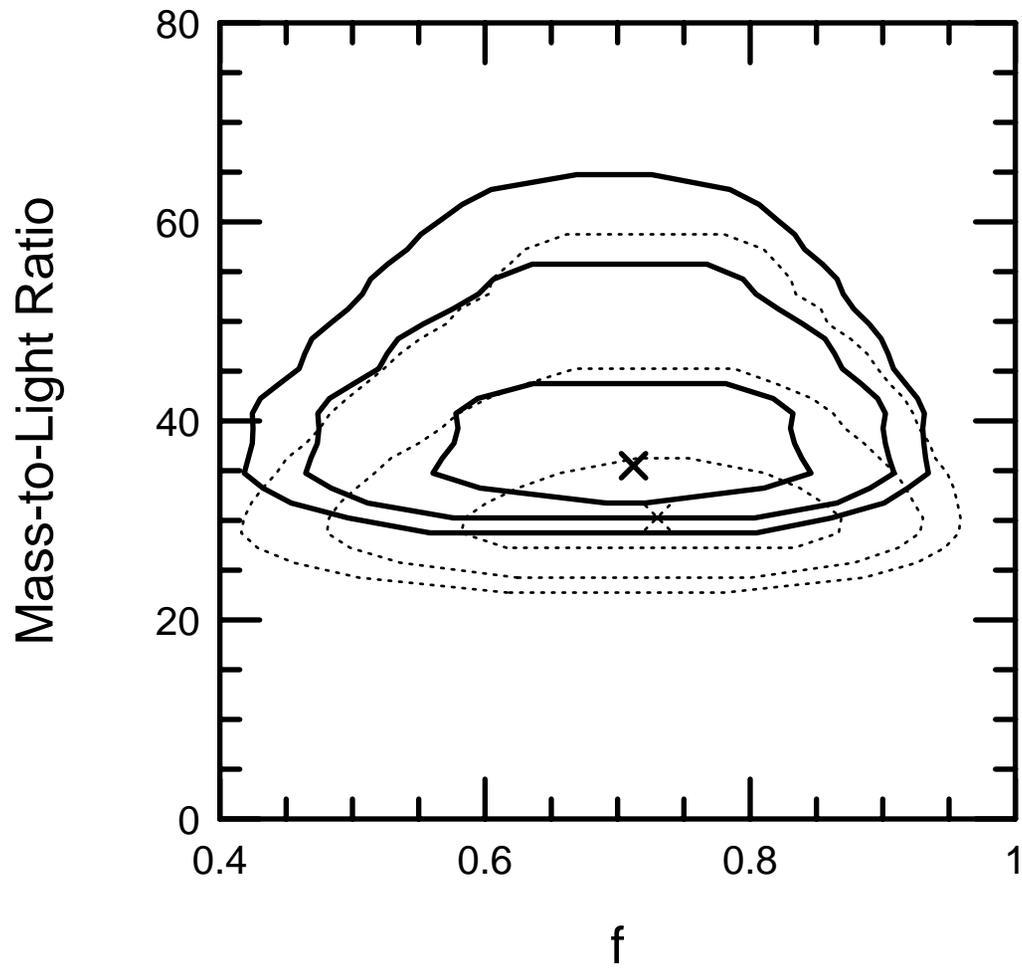
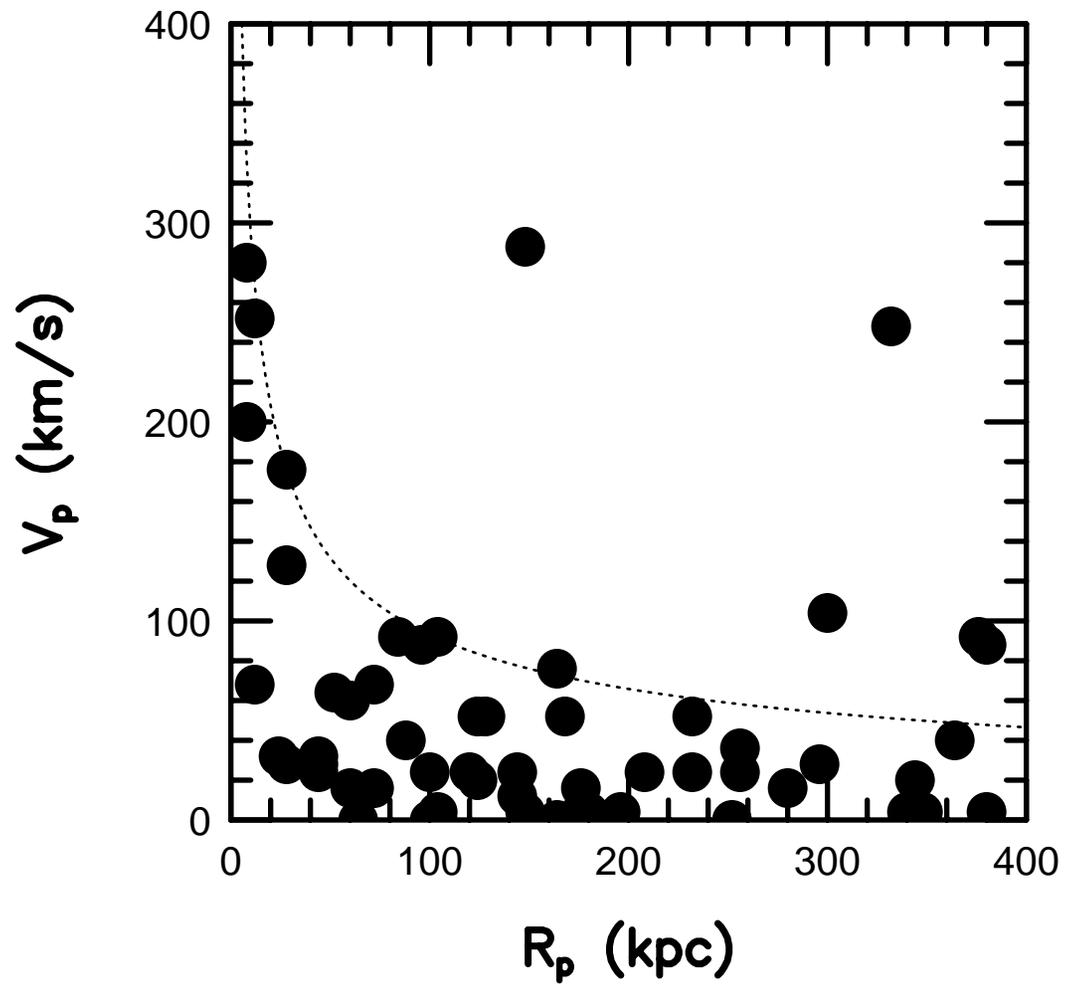
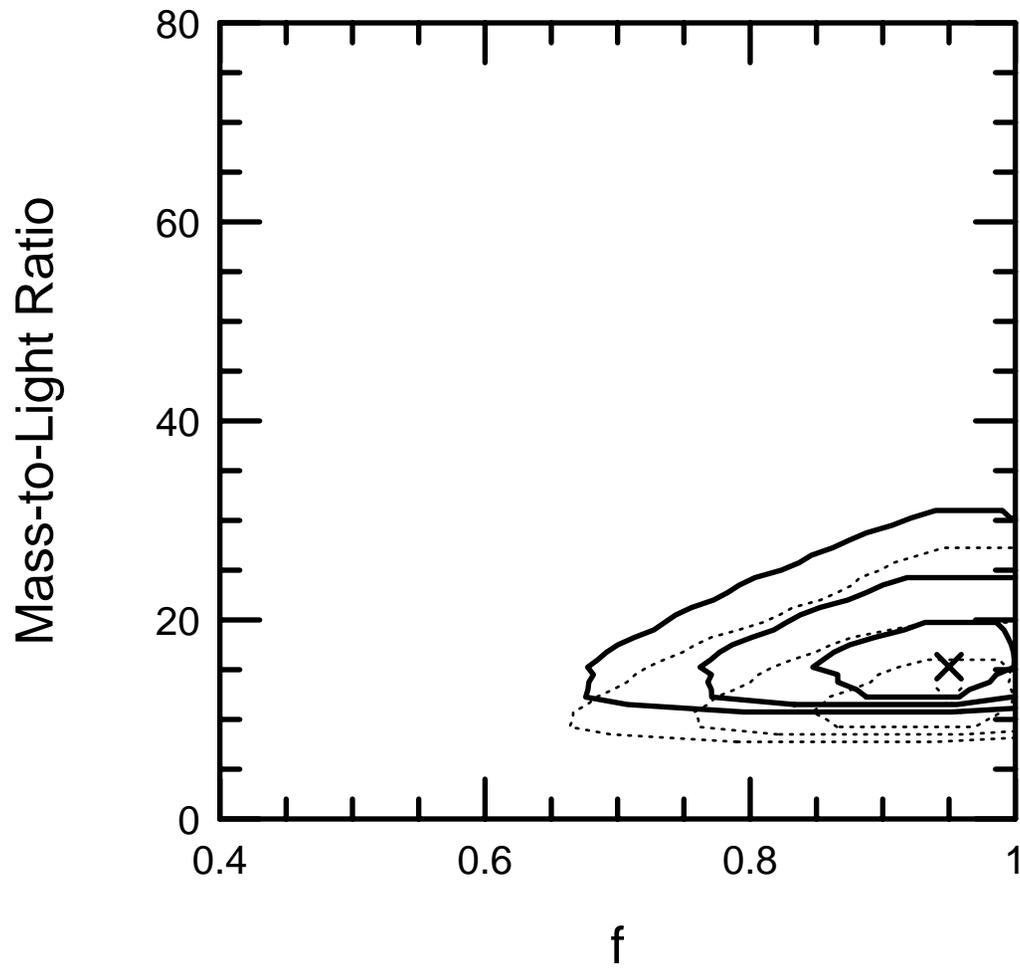
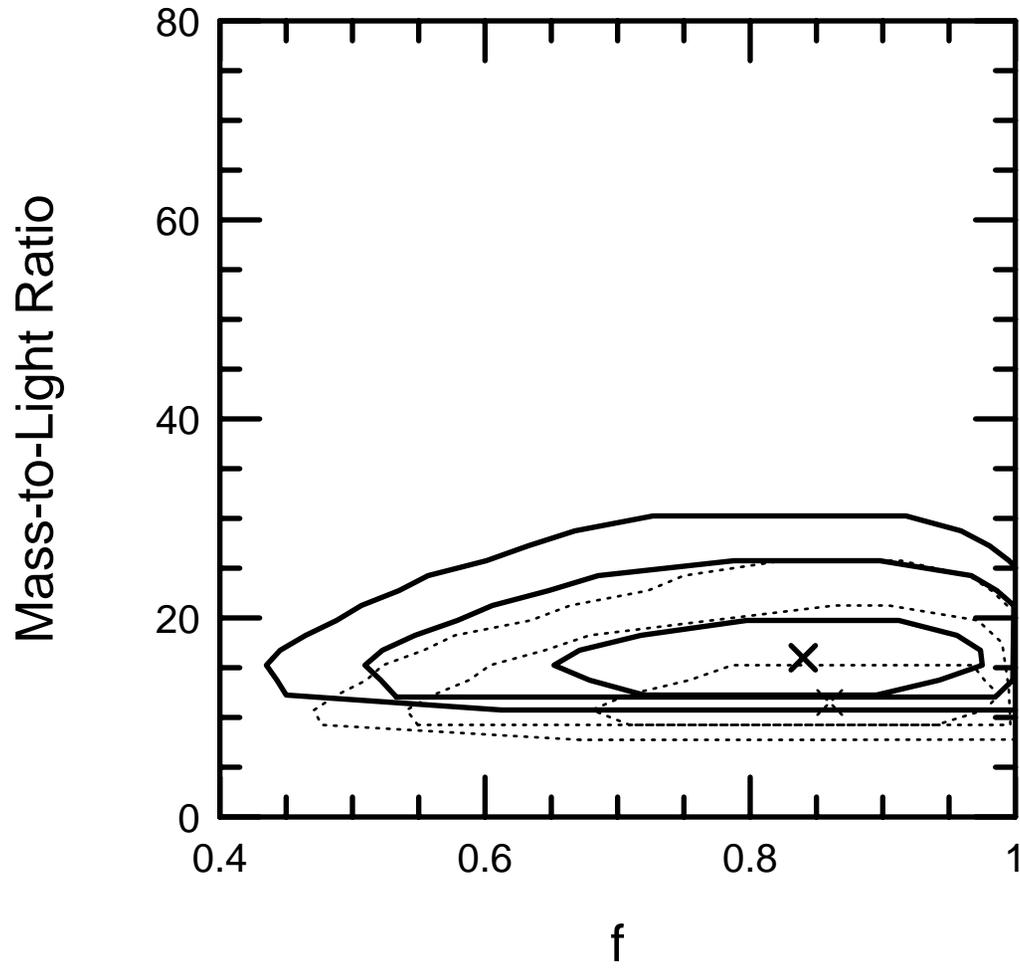
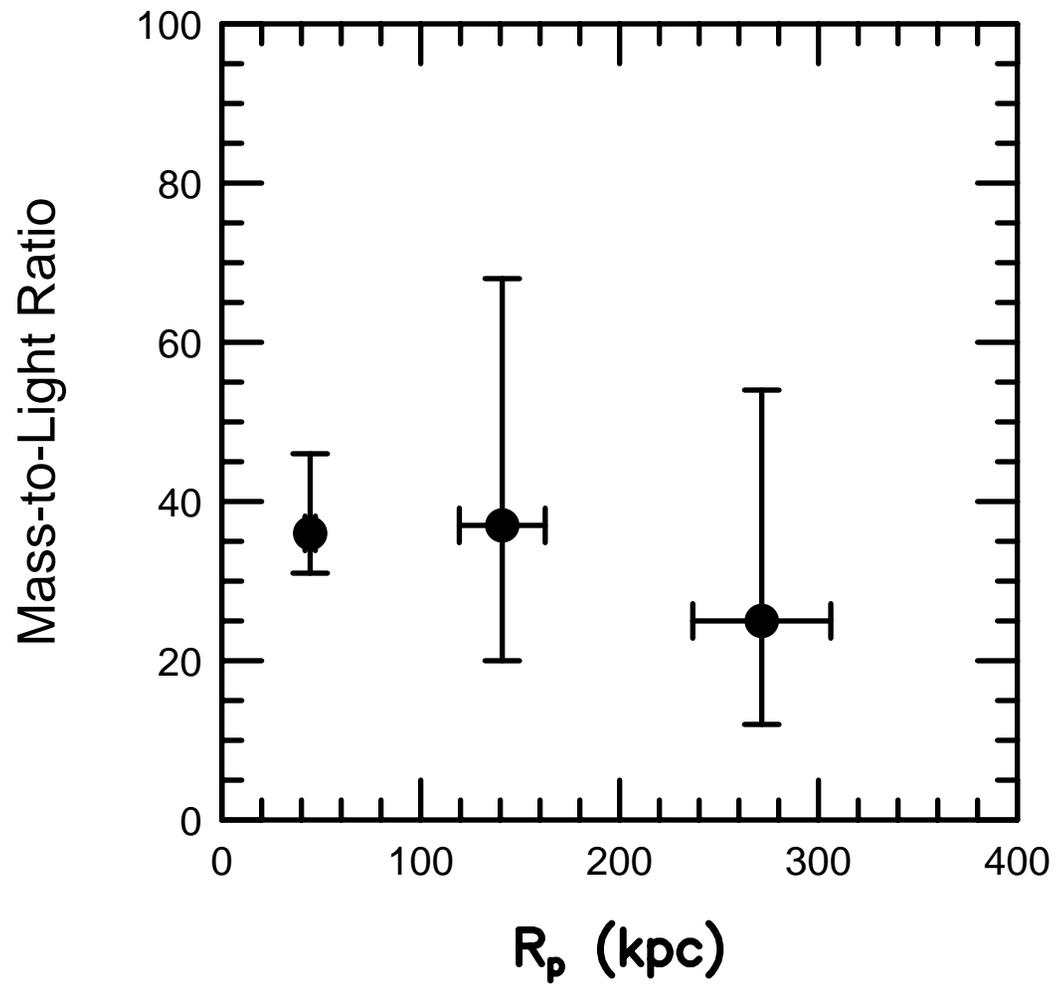
\end{document}